\pdfoutput=1
\RequirePackage{ifpdf}
\ifpdf 
\documentclass[pdftex]{sigma}
\else
\documentclass{sigma}
\fi

\numberwithin{equation}{section}

\def\II{\hbox{{1}\kern-.25em\hbox{l}}}

\begin{document}

\allowdisplaybreaks

\newcommand{\arXivNumber}{1711.07822}

\renewcommand{\thefootnote}{}

\renewcommand{\PaperNumber}{030}

\FirstPageHeading

\ShortArticleName{$\mathrm{SL}(2,\mathbb{C})$ Gustafson Integrals}

\ArticleName{$\boldsymbol{\mathrm{SL}(2,\mathbb{C})}$ Gustafson Integrals\footnote{This paper is a~contribution to the Special Issue on Elliptic Hypergeometric Functions and Their Applications. The full collection is available at \href{https://www.emis.de/journals/SIGMA/EHF2017.html}{https://www.emis.de/journals/SIGMA/EHF2017.html}}}

\Author{Sergey \'E.~DERKACHOV~$^\dag$, Alexander N.~MANASHOV~$^{\dag\ddag\S}$ and Pavel A.~VALINEVICH~$^\dag$}

\AuthorNameForHeading{S.\'E.~Derkachov, A.N.~Manashov and P.A.~Valinevich}

\Address{$^\dag$~Saint-Petersburg Department of Steklov Mathematical Institute\\
\hphantom{$^\dag$}~of Russian Academy of Sciences, Fontanka 27, 191023 St.~Petersburg, Russia}
\EmailD{\href{mailto:derkach@pdmi.ras.ru}{derkach@pdmi.ras.ru}, \href{mailto:valinevich@pdmi.ras.ru}{valinevich@pdmi.ras.ru}}

\Address{$^\ddag$~Institut f\"ur Theoretische Physik, Universit\"at Hamburg, D-22761 Hamburg, Germany}
\EmailD{\href{mailto:alexander.manashov@desy.de}{alexander.manashov@desy.de}}

\Address{$^\S$~Institute for Theoretical Physics, University of Regensburg, D-93040 Regensburg, Germany}

\ArticleDates{Received January 19, 2018, in final form March 24, 2018; Published online March 31, 2018}

\Abstract{It was shown recently that many of the Gustafson integrals appear in studies of the $\mathrm{SL}(2,\mathbb{R})$ spin chain models. One can hope to obtain a generalization of the Gustafson integrals considering spin chain models with a different symmetry group. In this paper we analyse the spin magnet with the $\mathrm{SL}(2,\mathbb{C})$ symmetry group in case of open and periodic boundary conditions and derive several new integrals.}


\Classification{81R12; 17B80; 33C70}

\renewcommand{\thefootnote}{\arabic{footnote}}
\setcounter{footnote}{0}

\section{Introduction}\label{sect:intro}

Gustafson's integrals are the multidimensional generalization of the classical Mellin--Barnes integrals. These integrals
\cite{Gustafson92,Gustafson94,Gustafson94-2} together with their $q$-deformed and elliptic analogs~\cite{Forrester-Warnaar,Spiridonov1,Spiridonov:2009za,Spiridonov06,Stockman} play an important role in the theory of special functions of many variables, the theory of random matrices, applied mathematics, etc. Many of these integrals have a group theoretic interpretation. Namely, it was shown in~\cite{Gustafson94,Gustafson94-2} that they can be represented as integrals over the corresponding compact simple Lie groups. At the same time it was known that some of these integrals appear naturally in the theory of integrable models. For instance, the classical Barnes first lemma follows from the Yang--Baxter relation for the $\mathcal{R}$ matrix for Toda chain. It turns out that such relations are not accidental. It is shown in~\cite{Derkachov:2016dhc} that many of the Gustafson integrals arise from certain relations between matrix elements in the $\text{SL}(2,\mathbb{R})$ spin chain models.

The spin chain models give an example of nontrivial integrable systems which can be solved by the quantum inverse scattering method (QISM)~\cite{MR1616371, Faddeev:1979gh,Kulish:1981gi,Kulish:1981bi,Sklyanin:1988yz,Sklyanin:1991ss}. One of the techniques available in the QISM is the separation of variables (SoV) method developed by E.K.~Sklya\-nin~\cite{Sklyanin:1991ss}. He has shown that the system of eigenfunctions of the elements of a monodromy matrix provides a convenient basis for studies of spin chain models. The SoV method is es\-pe\-cial\-ly suitable for the system with an infinite-dimensional Hilbert space such as spin chain models with noncompact symmetry groups, like $\text{SL}(2,\mathbb{R})$ or $\text{SL}(2,\mathbb{C})$. The eigenfunctions of the monodromy matrix elements for these models can be constructed explicitly in the form of multidimensional integrals which have a remarkably simple iterative structure~\cite{Belitsky:2014rba,Derkachov:2001yn,Derkachov:2002tf,Derkachov:2003qb,Derkachov:2014gya}. Moreover, these eigenfunctions can be interpreted as Feynman integrals of a certain type that facilitates calculations of scalar products between different eigenfunctions. As a rule such scalar products are given by a pro\-duct of the $\Gamma$-functions depending on the separated variables~-- parameters which characterize the eigenfunction. Using the com\-ple\-te\-ness of the corresponding bases and re-expanding one set of the eigenfunctions over the other set, one arrives at integral identities which are, in the case of the $\text{SL}(2,\mathbb{R})$ spin chains, easy to identify as Gustafson's integrals~\cite{Derkachov:2016dhc}. Moreover, proceeding along these lines it was possible to derive several new integrals.

It would be logical to extend this approach to spin chains with a different symmetry group. The most obvious candidate is the group $\text{SL}(2,\mathbb{C})$. The first step in this direction was done in~\cite{Derkachov:2016ucn} where a~generalization of the first Gustafson integral was obtained by analysing the closed $\text{SL}(2,\mathbb{C})$ spin magnet. It is remarkable enough that being written in terms of $\Gamma$-functions associated with the complex field $\mathbb{C}$~\cite{GelfandGraevRetakh04} the corresponding integral identity preserves its functional form. In this paper we derive the $\text{SL}(2,\mathbb{C})$ counterparts for another two Gustafson's integrals (the integrals~(2) and~(3) in \cite{Derkachov:2016dhc}). To this end we construct the SoV representation for the open spin chain magnets and calculate the transition matrix between the SoV representation for the {\it closed} and {\it open} spin chains. Using the completeness of the SoV representations we derive $\text{SL}(2,\mathbb{C})$ version of two integral identities that presents the main result of this paper.

Let us note that similar identities which involve integration and summation were considered in~\cite{Bazhanov:2013bh,Bazhanov:2007vg,Bazhanov:2010kz,Kels:2013ola,Kels:2015bda}. They are obtained as a limiting case of the elliptic star-triangle relation~\cite{Bazhanov:2010kz,Spiridonov:2010em}. The integrals derived in this paper also are obtained by a rather indirect approach. In our opinion it would be desirable to have a more direct proof of these identities using the techniques developed in~\cite{Gustafson94,Gustafson94-2,Spiridonov1,Spiridonov06}.

The paper is organized as follows: In Section~\ref{sect:magnets} we briefly review the formulation of the $\mathrm{SL}(2,\mathbb{C})$ spin magnets and the SoV construction. Section~\ref{sect:products} contains details of the calculation of relevant scalar products. The new integral identities are given in Section~\ref{sect:identities}. Sec\-tion~\ref{sect:summary} contains our summary and elements of the diagrammatic technique are collected in Appendix~\ref{sect:Diagram}.

\section[$\mathrm{SL}(2,\mathbb{C})$ spin magnets]{$\boldsymbol{\mathrm{SL}(2,\mathbb{C})}$ spin magnets}\label{sect:magnets}

The quantum ${\rm SL}(2,\mathbb{C})$ spin magnet is a one di\-mens\-ional lattice system which generalizes the well-known $\mathrm{XXX}_{s}$ chain to the case of complex spins. The dy\-na\-mi\-cal variables are two copies of spin generators $S^{(k)}_\alpha$, $\bar S^{(k)}_\alpha$, defined at each site, $k=1,\ldots,N$. Here and below $N$ is the length of the spin chain. The generators at the same site satisfy the standard ${\mathfrak{sl}}(2)$ commutation relation
\begin{alignat*}{3}
& [S_+,S_-]=2S_0 ,\qquad && [S_0 ,S_\pm]=\pm S_\pm , & \\
& [\bar S_+,\bar S_-]=2\bar S_0 , \qquad && [\bar S_0 ,\bar S_\pm]=\pm \bar S_\pm .&
\end{alignat*}
The generators at different sites commutes. All commutators between holomorphic opera\-tors~($S_\alpha$) and anti-holomorphic $(\bar S_\alpha$) vanish. The only interaction between two sectors comes from the conditions imposed on the wave functions.

We assume that the generators belong to a uni\-tary continuous principal series representation of the ${\rm SL}(2,\mathbb{C})$ group. Such a~representation is de\-ter\-mi\-ned by two complex spins, $s$ and $\bar s$, which are parameterized by a (half)integer number $n$ and a real number $\nu$~\cite{Gelfand}
\begin{gather}\label{spin-nnu}
s= {1/2+n/2}+i\nu, \qquad \bar s= {1/2-n/2}+i\nu.
\end{gather}
In the standard realization the generators are first-order differential operators
\begin{alignat*}{4}
& S_- =-\partial_z,\qquad && S_0 =z\partial_z+s, \qquad && S_+=z^2\partial_z+2s z ,& \\
& \bar S_- =-\partial_{\bar z}, \qquad && \bar S_0 ={\bar z}\partial_{\bar z}+\bar s, \qquad && \bar S_+=\bar z^2\partial_{\bar z}+2\bar s \bar z,&
\end{alignat*}
which are adjoint to each other, $S_\alpha^\dagger = -\bar S_\alpha,$ with respect to the $L_2(\mathbb{C})$ scalar product
\begin{gather*}
( \Phi | \Psi) = \int {\rm d}^2z \overline{\Phi(z,\bar z)} \Psi(z,\bar z) .
\end{gather*}
The Hilbert space of the model is given by a direct product of $N$ copies of the $L_2(\mathbb{C})$ spaces,
\begin{gather}\label{Hilbert-space}
\mathbb{H}_N=\mathbb{V}_1\otimes \mathbb{V}_2\otimes\cdots\otimes \mathbb{V}_N, \qquad \mathbb{V}_k=L_2(\mathbb{C}) .
\end{gather}
We will consider only the homogeneous chains, i.e., $s_1=s_2=\cdots=s_N=s$. Finally, since any equation in holomorphic sector has its anti-holomorphic twin copy we will write only the holomorphic version in most cases.

\subsection{SoV for closed spin chain}

The QISM allows one to construct a family of non\-trivial commuting operators acting on the Hilbert space of the model. The fundamental object in this approach is the so-called Lax operator defined as
\begin{gather*}
L(u) =u +i(\sigma\otimes S) , \qquad \bar L(\bar u) =\bar u +i\big(\sigma\otimes \bar S\big) ,
\end{gather*}
where $\sigma^\alpha$ is the Pauli matrices, {$(\sigma\otimes S)=\sum_\alpha\sigma^\alpha S_\alpha$} and complex numbers $u$, $\bar u$ are called the spectral parameters. Taking a product of Lax operators one gets a mono\-dromy matrix for a~closed spin chain
\begin{gather}\label{monodromy}
T_N(u)=L_1(u) \cdots L_N(u)=
\begin{pmatrix}
A_N(u)& B_N(u)\\
C_N(u)& D_N(u)
\end{pmatrix}.
\end{gather}
Replacing $L_k(u)\to\bar L_k(\bar u)$ one gets the anti-holo\-mo\-rphic monodromy matrix, $\overline T_N(\bar u)$. It has to be stressed here that we do not assume any relation between the parameters $u$ and $\bar u$. It is shown in the QISM~\cite{Faddeev:1979gh} that the entries of the monodromy matrix, which are polynomials in a~spectral parameter, form commuting operator families, i.e.,
\begin{gather*}
[A_N(u), A_N(v)]=0, \qquad [B_N(u), B_N(v)]=0 ,
\end{gather*}
etc. Commutativity implies that the eigenfunctions of the operators $A_N$, $\bar A_N$ ($B_N$, $\bar B_N$) do not depend on the spectral parameters. They form a basis of Sklyanin's representation of the separated variables~\cite{Sklyanin:1991ss} and were constructed in the explicit form in~\cite{Derkachov:2001yn,Derkachov:2014gya}. One can find a~detailed derivation in~\cite{Derkachov:2001yn,Derkachov:2014gya,Derkachov:2016ucn}. We give only final expressions for the eigenfunctions which match exactly those in~\cite{Derkachov:2016ucn} including the normalization factors. The eigenfunction $\Psi_A ({\boldsymbol{x}}|\boldsymbol{z})$ satisfies the equations
\begin{gather*}
A_N(u) \Psi_A ({\boldsymbol{x}}|\boldsymbol{z})= \prod_{k=1}^N (u-x_k)\Psi_A ({\boldsymbol{x}}|\boldsymbol{z}) ,\\
\bar A_N(\bar u) \Psi_A({\boldsymbol{x}}|\boldsymbol{z}) = \prod_{k=1}^N (\bar u-\bar x_k) \Psi_A ({\boldsymbol{x}}|\boldsymbol{z}) ,
\end{gather*}
where $x_k$ ($\bar x_k=x_k^*$) are the separated variables which take the form
\begin{gather}\label{xk}
x_k=-in_k/2 +\nu_k
\end{gather}
with $n_k-n\in \mathbb{Z}$ and $\nu_k\in \mathbb{R}$. We use boldface letters for sets, $\boldsymbol{x}=\{x_1,\ldots,x_N\}$, etc.
\begin{figure}[t]\centering
\includegraphics[width=0.4\linewidth]{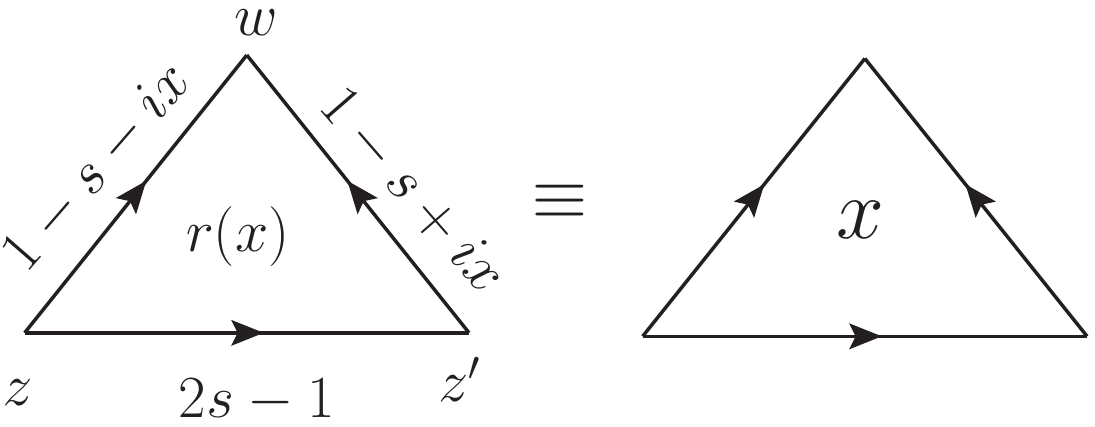}
\caption{The diagrammatic representation for the function $\chi_x(w,z,z')$. The arrow from the point $z$ to $w$ with the index $\alpha$ stands for $[w-z]^{-\alpha}$.}\label{fig:def}
\end{figure}

Similarly, the eigenfunctions of the operator $B_N$ obey the equations
\begin{gather*}
\begin{split}&
B_N(u) \Psi_B (p,{\boldsymbol{x}}|\boldsymbol{z}) = p \prod_{k=1}^{N-1} (u-x_k) \Psi_B (p,{\boldsymbol{x}}|\boldsymbol{z}) ,\\
& \bar B_N(\bar u) \Psi_B (p,{\boldsymbol{x}}|\boldsymbol{z})= \bar p\prod_{k=1}^{N-1} (\bar u-\bar x_k)\Psi_B (p,{\boldsymbol{x}}|\boldsymbol{z}) ,
\end{split}
\end{gather*}
where $\bar p =p^\ast$ and $\boldsymbol{x}=\{x_1,\ldots,x_{N-1}\}$ and $x_k$ takes the form~\eqref{xk}. The eigenfunctions can be written as
\begin{gather}
\Psi_B (p ,{\boldsymbol{x}}|\boldsymbol{z}) = |p|^{N-1}\Lambda_{N}(x_1)\cdots \Lambda_{2}(x_{N-1}) \mathrm{e}^{i p z_1+i \bar{p} \bar{z}_1},\notag\\
\label{reprPsiA} \Psi_A ({\boldsymbol{x}}|\boldsymbol{z}) = \widetilde \Lambda_{N}(x_1)\cdots \widetilde \Lambda_{2}(x_{N-1})
\widetilde \Lambda_{1}(x_{N}) .
\end{gather}
The layer operators $\Lambda_k(x)$ and $\widetilde \Lambda_k(x)$ map functions of $k-1$ variables to functions of $k$ variables. Namely,
\begin{gather*}
\big[\Lambda_k(x) \Phi\big](\boldsymbol{z})= \prod_{i=1}^{k-1} \int {\rm d}^2 w_i \chi_x(z_i,z_{i+1},w_i) \Phi(\boldsymbol{w}) ,
\end{gather*}
where $\boldsymbol{z}=\{z_1,\ldots,z_k\}$ and $\boldsymbol{w}=\{w_1,\ldots,w_{k-1}\}$. The function $\chi_x$ takes the form (see also Fig.~\ref{fig:def})
\begin{gather*}
\chi_x(z,z',w) =r(x) [z'-z]^{1-2s} [w-z]^{s+ix-1}[w-z']^{s-ix-1},
\end{gather*}
where $[z]^{\alpha}\equiv z^\alpha \bar z^{\bar\alpha}$, the normalization factor is
\begin{gather*}
r(x)=a(s+ix) a(\bar s-i\bar x)
\end{gather*}
 and the function $a(\alpha)$ is defined as follows
 \begin{gather*}
a(\alpha)=\Gamma(1-\bar\alpha)/\Gamma(\alpha).
 \end{gather*}
The second layer operator $\widetilde \Lambda_k(x)$ reads
\begin{gather*}
\widetilde\Lambda_k(x) \equiv [z_k]^{-s+ix} \Lambda_k(x) .
\end{gather*}
Let us note that for the chosen normalization the layer operators satisfy the exchange relations
\begin{gather*}
\Lambda_k(x)\Lambda_{k-1}(x')=\Lambda_k(x')\Lambda_{k-1}(x)
\end{gather*}
and the same for $\widetilde \Lambda_k$. This ensures that the functions $\Psi_B(p,\boldsymbol{x})$ and $\Psi_A(\boldsymbol{x})$ are symmetric functions of the separated variables.

\begin{figure}[t]\centering
\includegraphics[width=0.45\linewidth]{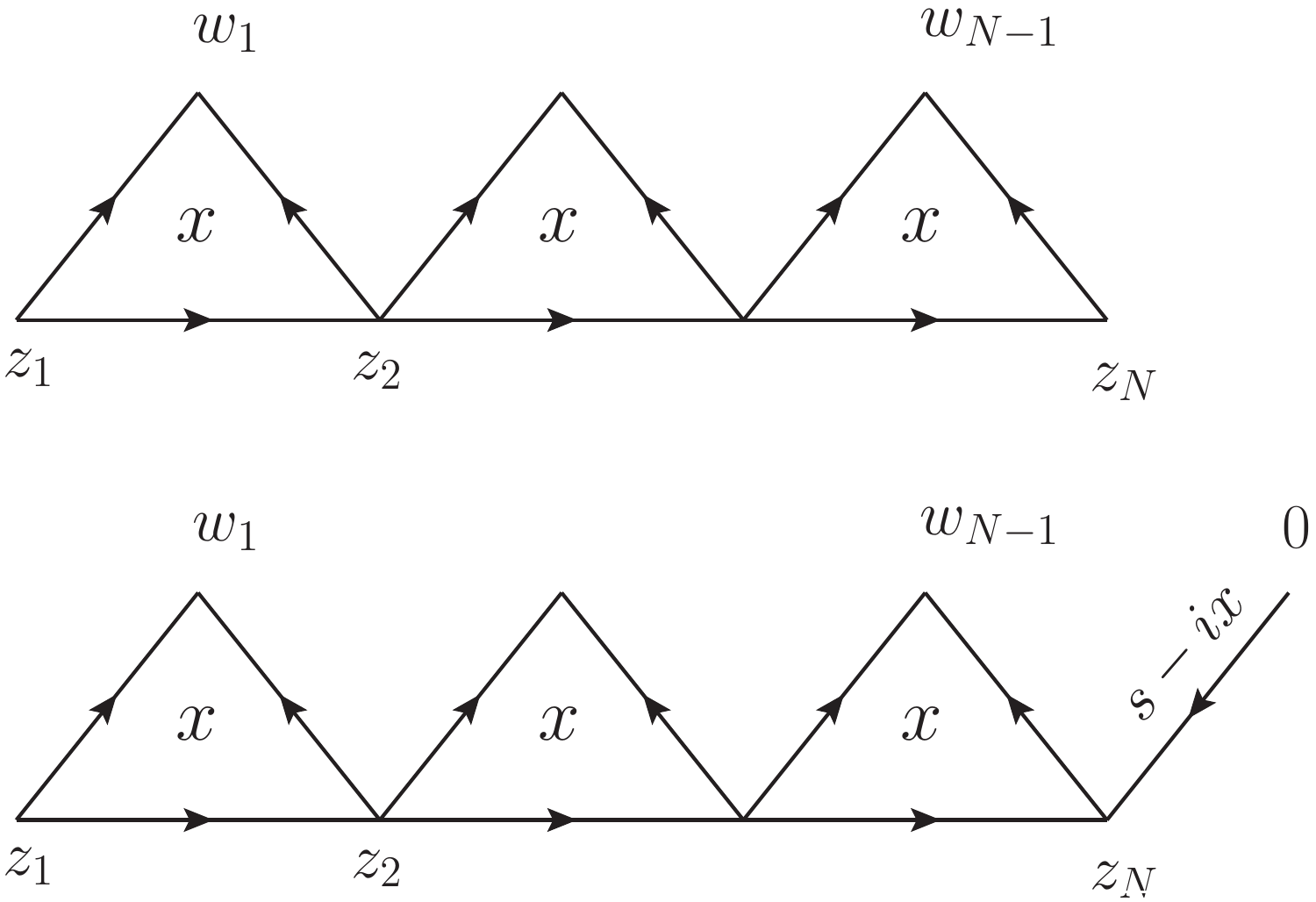}
\caption{The diagrammatic representation for the kernels of the layer operators $\Lambda_N(x)$ (upper diagram) and $\widetilde \Lambda_N(x)$ (bottom diagram) for $N=4$.}\label{fig:Lambda}
\end{figure}

The kernels of the operators and the eigenfunctions can be represented in the form of Feynman diagrams. This appears to be quite useful for proving certain operator identities. The diagrammatic representation of the layer operators $\Lambda_N$ and $\widetilde \Lambda_N$ is shown in Fig.~\ref{fig:Lambda}.

Being the eigenfunctions of the self-adjoint operators these functions form a complete ortho\-go\-nal basis in the Hilbert space~$\mathbb{H}_N$ (equation~\eqref{Hilbert-space}),
\begin{gather*}
\big( \Psi_A({\boldsymbol{x}'}), \Psi_A({\boldsymbol{x})}\big)= \big(\boldsymbol{ \mu}^{(A)}_N(\boldsymbol{x})\big)^{-1} \delta_N(\boldsymbol{x}-\boldsymbol{x}') ,
\end{gather*}
and
\begin{gather*}
\big({\Psi_B(p',{\boldsymbol{x}'})}, \Psi_{B} (p,{\boldsymbol{x}})\big) = \big(\boldsymbol{\mu}^{(B)}_N(\boldsymbol{x})\big)^{-1} \delta^2(\vec{p}-\vec{p}\,{}^\prime) \delta_{N-1}(\boldsymbol{x}-\boldsymbol{x}') .
\end{gather*}
Here
\begin{gather*}
\delta_N(\boldsymbol{x}-\boldsymbol{x}')=\frac1{N!} \sum_{s\in S_N}\prod_{k=1}^N\delta^{(2)}\big(x_k-x'_{s(k)}\big),
\end{gather*}
where the sum goes over all permutations of $N$ ele\-ments and $\delta^{(2)}(x-x')\equiv \delta_{n n'} \delta(\nu-\nu')$. The weight functions
$\boldsymbol{ \mu}^{(A,B)}_N(\boldsymbol{x})$ (Sklyanin's measures) take the form~\cite{Derkachov:2001yn,Derkachov:2014gya}
\begin{gather*}
\boldsymbol{\mu}^{(B)}_N(\boldsymbol{x})=\frac1{(N-1)!} \frac{2\pi^{-N^2}}{(2\pi)^{N}}\prod_{1\leq k<j\leq N-1} [x_k-x_j],\\
\boldsymbol{\mu}^{(A)}_N(\boldsymbol{x}) =\frac1{N!}\frac{\pi^{-N^2}}{(2\pi)^{N}} \prod_{1\leq k<j\leq N}[x_k-x_j],
\end{gather*}
where we introduced the notation $[x]\equiv x\bar x$.

A completeness condition for the $B$-system reads
\begin{gather*}
\prod_{k=1}^{N}\delta^{(2)}(\vec{z}_k-\vec{z}_k^{\,\prime}) =\int {\rm d}^2 p \int \mathcal{D}_{N-1} \boldsymbol{x} \boldsymbol{\mu}^{(B)}_N(\boldsymbol{x}) \Psi_B({p,\boldsymbol{x}}|\boldsymbol{z})\overline{\Psi_B({p,\boldsymbol{x}}|\boldsymbol{z}')} ,
\end{gather*}
where the integration measure is defined as
\begin{gather*}
\int \mathcal{D}_{N-1}\boldsymbol{x}=\prod_{k=1}^{N-1} \left(\sum_{n_k=-\infty}^{\infty}\int_{-\infty}^{\infty} {\rm d}\nu_k\right).
\end{gather*}
The sum goes over integer or half-integer $n$ depending on whether $[s]$ is an integer or half-integer. A similar expression for the $A$-system can be found in~\cite{Derkachov:2016ucn}.

We also write down the expressions for two matrix elements obtained in~\cite{Derkachov:2016ucn}. First of them is the matrix ele\-ment of the shift operator,
\begin{gather*}
T_{z_0} \Phi(\boldsymbol{z})=\Phi(z_1-z_0,\ldots,z_N-z_0),
\end{gather*}
between $A$ states
\begin{gather}\label{Txx-result}
T_{z_0}(\boldsymbol{x},\boldsymbol{x}^\prime)=\big( \Psi_A(\boldsymbol{x}'),T_{z_0}\Psi_A(\boldsymbol{x})\big) =(-1)^{[A_X]}[z_0]^{i(A_{X'}-A_X)}\prod_{k,j=1}^{N} q(x_k,x'_j) .
\end{gather}
Here $(-1)^{[A]}\equiv(-1)^{A-\bar A}$, $ A_X =\sum_k (s-ix_k) $ and
\begin{gather}\label{q-f-def}
q(x,x')=\pi a\big(1+i(x-x')\big) \frac{a(\bar s- i\bar x)}{a(s-ix')} .
\end{gather}
The second one is a scalar product
\begin{gather*}
\big( \Psi_B(p,\boldsymbol{x}'),\Psi_A(\boldsymbol{x})\big) = {|p|^{-N-1}} [p]^{A_X} S_{BA}(\boldsymbol x,\boldsymbol x'),
\end{gather*}
where
\begin{gather*}
S_{BA}(\boldsymbol x,\boldsymbol x')=\pi^{N} i^{[A_X]} \prod_{k=1}^{N}a(\bar s-i\bar x_k) \prod_{j=1}^{N-1}q(x_k,x'_j).
\end{gather*}
Generalization of the first Gustafson integral to the complex case follows from the re-expansion of the matrix element~\eqref{Txx-result} over the $B$-system. To proceed fur\-ther we have to construct the SoV representation for the open $\text{SL}(2,\mathbb{C})$ spin chain.

\subsection{SoV for open spin chain}

\begin{figure}[t]\centering
\includegraphics[width=0.5\linewidth]{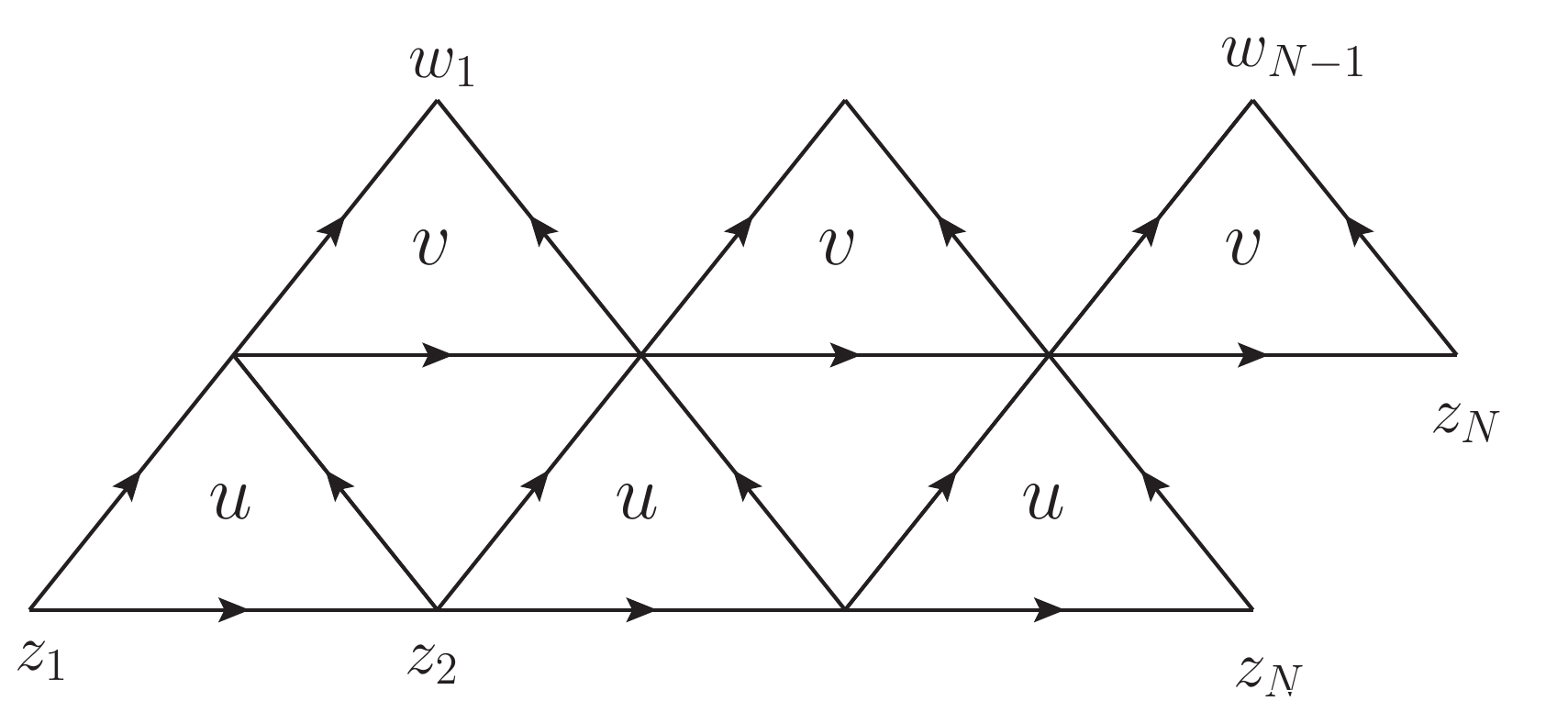}
\caption{The diagrammatic representation for the function $G_N(u,v|\boldsymbol{z},\boldsymbol{w})$ for $N=4$.}\label{fig:Guv}
\end{figure}

A systematic approach for constructing the integrable models with nontrivial boundary conditions has been developed by E.K.~Sklyanin~\cite{Sklyanin:1988yz}. The monodromy matrix for an open spin chain is given by the following expres\-sion\footnote{This definition differs from a standard one, $T_N(u) T_N^{-1}(-u+i)$, by a numerical factor.}
\begin{gather*}
\mathbb{T}_N(u)= T_N(-u) \sigma_2 T^{t}_N(u) \sigma_2 =
\begin{pmatrix}
\mathbb{A}_N(u) & \mathbb{B}_N(u)\\
\mathbb{C}_N(u) & \mathbb{D}_N(u)
\end{pmatrix} ,
\end{gather*}
where $T_N(u)$ is the monodromy matrix for the closed spin chain, equation~\eqref{monodromy}. It is shown in the QISM that the operators $\mathbb{B}_N(u)$ form a commuting family, $[\mathbb{B}_N(u),\mathbb{B}_N(v)]=0$, and therefore can be dia\-go\-na\-li\-zed simultaneously. By a construction $\mathbb{B}_N(u)$ is a poly\-no\-mial of the degree $2N-1$ in $u$. It can be shown, see, e.g., \cite{Derkachov:2003qb}, that it vanishes at $u=-i/2$ and satisfies the equation{\samepage
\begin{gather*}
{(-2u+i)} {\mathbb{B}_N(u)}=(2u+i) {\mathbb{B}_N(-u)} .
\end{gather*}
Thus the operator $\widehat{ \mathbb{B}}_N(u)={ \mathbb{B}}_N(u)/(2u+i)$ is a polynomial of the degree $N-1$ in $u^2$.}

In order to construct the ei\-gen\-functions of $\widehat{ \mathbb{B}}_N(u)$ we follow the approach of~\cite{Derkachov:2003qb}. Let us define a function $G_N(u,v|\boldsymbol{z},\boldsymbol{w})$ as follows
\begin{gather*}
G_N(u,v|\boldsymbol{z},\boldsymbol{w}) =\prod_{k=1}^{N-1}\int {\rm d}^2\xi_k \chi_u(z_k,z_{k+1},\xi_k) \chi_v(\xi_k,\xi_{k+1}, w_k)\Big|_{\xi_N=z_N}.
\end{gather*}
In the diagrammatic form this function is shown in Fig.~\ref{fig:Guv}. The function has certain properties that allow one to identify it with the kernel of the layer operator for the open spin chain. Namely, it can be verified with the help of the dia\-gram\-matic technique that $G_N(u,v)$ is invariant under the permutation $(u,\bar u)\leftrightarrow (v,\bar v)$. Moreover, one can check (see~\cite{Derkachov:2003qb}) that
\begin{gather}\label{annihilate}
\mathbb{B}_N(u)G_N(u,-u)= \bar{\mathbb{B}}_N(\bar u)G_N(u,-u)=0 .
\end{gather}
Therefore we define the layer operator for the open spin chain as follows
\begin{gather*}
\big[\boldsymbol\Lambda_k(x) \Phi\big](\boldsymbol{z})=\prod_{i=1}^{k-1}\int {\rm d}^2 w_i G_k(x,-x|\boldsymbol{z},\boldsymbol{w})\Phi(\boldsymbol{w}).
\end{gather*}
The layer operators are even functions of the spectral parameter $x$, $ \boldsymbol\Lambda_k(x)=\boldsymbol\Lambda_k(-x)$, and for the chosen normalization they satisfy the exchange relation
\begin{gather}\label{exchange-open}
\boldsymbol\Lambda_k(x)\boldsymbol\Lambda_{k-1}(x')=\boldsymbol\Lambda_k(x')\boldsymbol\Lambda_{k-1}(x).
\end{gather}
In order to prove such a relation it is sufficient to show that the kernels of operators on both sides of an equation coincide. It can be done diagrammatically with the help of the so-called integration rules given in Appendix~\ref{sect:Diagram}. (Detailed examples of an application of the diagrammatic technique can be found in~\cite{Derkachov:2001yn,Derkachov:2014gya}.)

The eigenfunction of the operator $\widehat{\mathbb{B}}_N$ takes exactly the same form as for the closed chain, the only modification being a replacement of the layer operators $\Lambda_k\mapsto\boldsymbol\Lambda_k$
\begin{gather}\label{PsiBopen}
\Psi_{\mathbb{B}}(p,\boldsymbol{x}|\boldsymbol{z})=|p|^{N-1}\boldsymbol\Lambda_N(x_1)\cdots \boldsymbol\Lambda_2(x_{N-1})e^{i p z + i\bar p \bar z}.
\end{gather}
Indeed, by virtue of equations~\eqref{exchange-open} and \eqref{annihilate} this function is annihilated by the operators $\widehat{\mathbb{B}}_N(\pm x_k)$, $\widehat{\overline{\mathbb{B}}}_N(\pm \bar x_k)$ and it is also an eigenstate of the mo\-men\-tum operator $\Big(\sum\limits_{k=1}^N S^{(k)}_- \Big) \Psi_{\mathbb{B}}(p,\boldsymbol{x})=p \Psi_{\mathbb{B}}(p,\boldsymbol{x})$. This implies that
\begin{gather*}
\widehat{\mathbb{B}}_N(u)\Psi_{\mathbb{B}}(p,\boldsymbol{x})= p\prod_{k=1}^{N-1}\big(u^2-x_k^2\big)\Psi_{\mathbb{B}}(p,\boldsymbol{x}) ,
\end{gather*}
and similar for $\widehat{\overline{\mathbb{B}}}_N(\bar u)$.

The single-valuedness requirement for the function~\eqref{PsiBopen} leads to quantization of the separated va\-ri\-ables: $s+ix_k -(\bar s+i\bar x_k)\in \mathbb{Z}$. It implies
\begin{gather*}
x_k=-\frac{in_k}2+\nu_k, \qquad \bar x_k=\frac{in_k}2+\nu_k,
\end{gather*}
where $n_k$ is a (half)integer such that $n_k-n\in \mathbb{Z}$ and $n$ is defined in~\eqref{spin-nnu}. The normalization condition for the eigenfunctions implies that $\nu_k\in \mathbb{R}$.

Since the function $\Psi_{\mathbb{B}}(p,\boldsymbol{x})$ is invariant under $x_k\to -x_k$ one can always assume that the separated variables belong to one of two regions: either $S_{\rm I} =\{\nu_k\in\mathbb{R}_+, n_k-n\in \mathbb{Z}\}$ or $S_{\rm II}=\{n_k\geq 0, \nu_k\in \mathbb{R}\}$.

\begin{figure}[t]\centering
\includegraphics[width=0.5\linewidth]{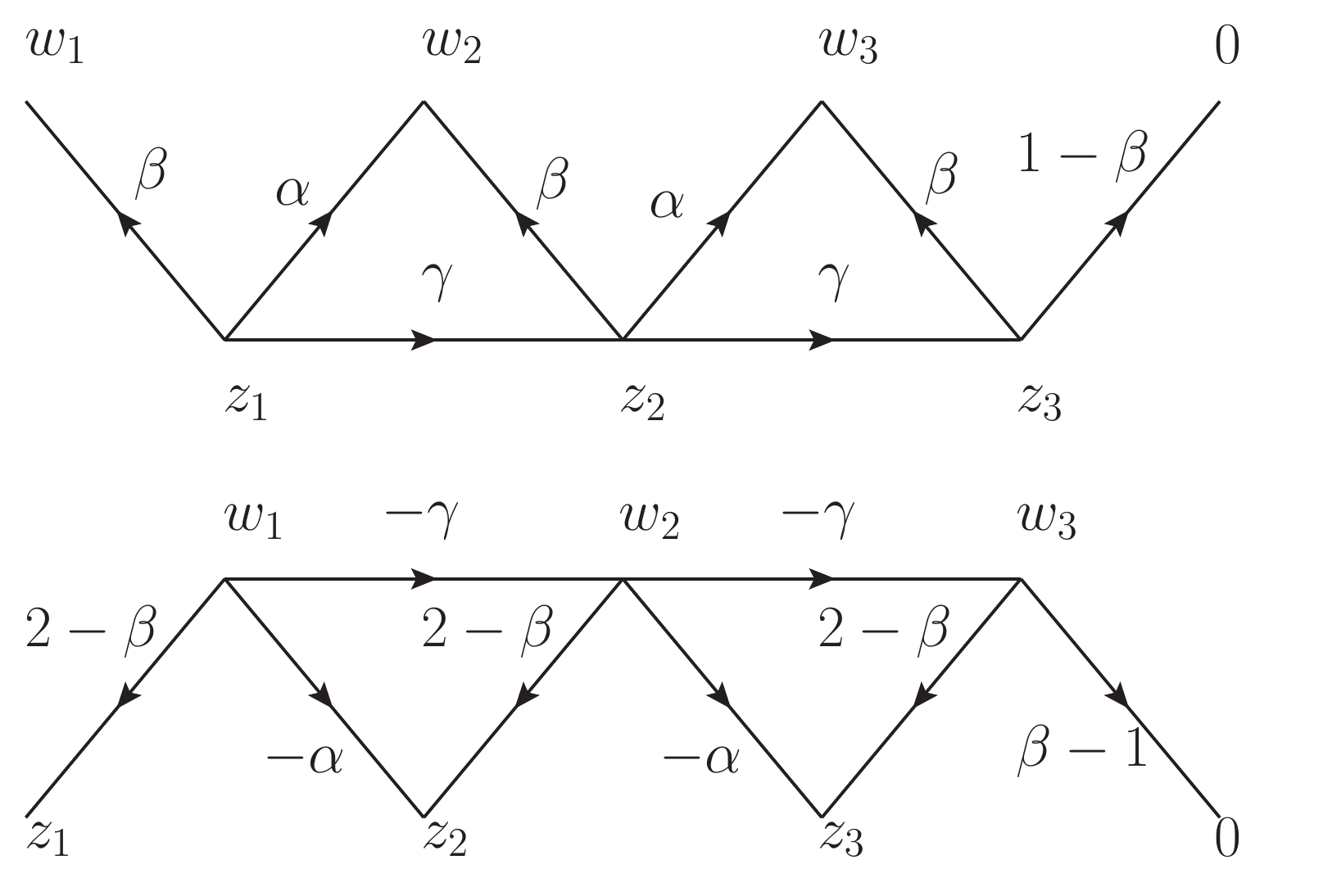}
\caption{The diagrammatic representation for the kernels of the operators $Q_{N}(x)$ (upper diagram) and $Q_N^{-1}(x)$ (bottom diagram) for $N=3$.
The bottom diagram has to be multiplied by a factor $\left(-(s-ix)(\bar s-i\bar x)/\pi^2\right)^{N}$. The indices take the following values: $\alpha=1-s-ix$, $\beta=1-s+ix$ and $\gamma=2s-1$.} \label{fig:QX}
\end{figure}

The calculation of the scalar product of eigenfunctions is based on the following relation for the layer operators
\begin{gather*}
\boldsymbol\Lambda^\dagger_k(x')\boldsymbol \Lambda_k(x) =\vartheta(x,x') \boldsymbol\Lambda_{k-1}(x) \boldsymbol\Lambda^\dagger_{k-1}(x') ,
\end{gather*}
which holds for $x\neq x'$ and where
\begin{gather*}
\vartheta(x,x')=\frac{\pi^2}{[x-x']}\frac{\pi^2}{[x+x']} .
\end{gather*}
Assuming that $\boldsymbol{x}\in S_I$ we obtain
\begin{gather*}
\big({\Psi_{\mathbb{B}}(p',{\boldsymbol{x}'})}, \Psi_{\mathbb{B}} (p,{\boldsymbol{x}})\big) = \big(\boldsymbol{\mu}^{(\mathbb{B})}_N(\boldsymbol{x})\big)^{-1} \delta^2(\vec{p}-\vec{p}\,{}^\prime) \delta_{N-1}(\boldsymbol{x}-\boldsymbol{x}') ,
\end{gather*}
where the weight function is given by
\begin{gather*}
\boldsymbol{\mu}^{(\mathbb{B})}_N(\boldsymbol{x})= \frac1{(N-1)!}\frac{\pi^{-2N^2}}{2^{N-1}}
\prod_{1\leq k<j\leq N-1}{[x_k-x_j]} \prod_{1\leq k\leq j\leq N-1} {[x_k+x_j]}.
\end{gather*}
A completeness condition for the $\mathbb{B}$-system reads
\begin{gather*}
\prod_{k=1}^{N}\delta^{(2)}(\vec{z}_k-\vec{z}^{\,\prime}_k) =\int {\rm d}^2 p \int_{S_I} \mathcal{D}_{N-1} \boldsymbol{x}
\boldsymbol{\mu}^{(\mathbb{B})}_N(\boldsymbol{x}) \Psi_{\mathbb{B}}({p,\boldsymbol{x}}|\boldsymbol{z})
\overline{\Psi_{\mathbb{B}}({p,\boldsymbol{x}}|\boldsymbol{z}')} ,
\end{gather*}
where
\begin{gather*}
\int_{S_I} \mathcal{D}_{N-1}\boldsymbol{x}=\prod_{k=1}^{N-1} \left(\sum_{n_k=-\infty}^{\infty}\int_{0}^{\infty} {\rm d}\nu_k\right)
\end{gather*}
and the sum goes over integers or half-integers if $n$ is an integer or half-integer, respectively.

\begin{figure}[t]\centering
\includegraphics[width=0.5\linewidth]{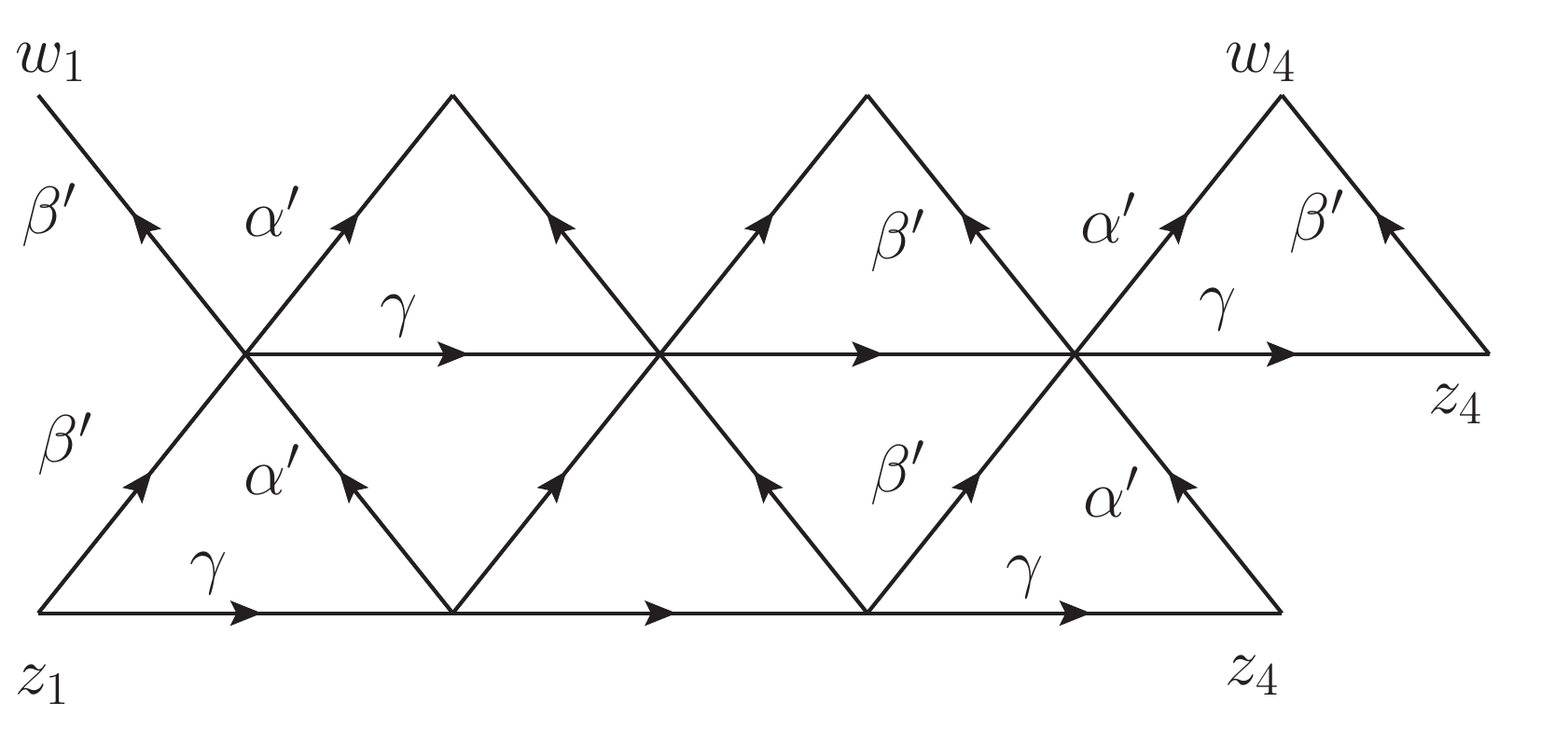}
\caption{The Baxter operator $\boldsymbol{Q}_{N}(x')$ for $N=4$. Here $\alpha'=1-s-ix'$, $\beta'=1-s+ix'$ and $\gamma=2s-1$. For $N=1$ the corresponding kernel takes form $\boldsymbol{Q}_{1}(x'|z,w)=[w-z]^{-1+s-ix'}$.} \label{fig:Blayer}
\end{figure}

\section{Scalar products}\label{sect:products}

In order to derive new integral identities we have to calculate the scalar product between the eigenfunctions of the closed and open spin chains. The calculation makes use of the recursive structure of the eigenfunctions. Namely, let us represent $\Psi_N^{(A)}(\boldsymbol{x}')$ and $\Psi_N^{(\mathbb{B})}({p,\boldsymbol{x}})$ as follows
\begin{gather*}
\Psi_N^{(A)}(\boldsymbol{x}')= \widetilde\Lambda_N(x'_1) \Psi_{N-1}^{(A)}(\boldsymbol{x}'_{N-1}) ,\\
\Psi_N^{(\mathbb{B})}({p,\boldsymbol{x}})=|p|\boldsymbol{\Lambda}_N(x_1)\Psi_{N-1}^{(\mathbb{B})}({p,\boldsymbol{x}_{N-1}}) .
\end{gather*}
Here $\boldsymbol{x}'_{N-1}\!=\!\{x'_2,\ldots,x'_{N}\}$ and similarly for $\boldsymbol{x}_{N-1}$. Then the scalar product $\big(\Psi_N^{(A)}(\boldsymbol{x}'),\Psi_N^{(\mathbb{B})}({p,\boldsymbol{x}})\big)$ can be cast into the form
\begin{gather*}
|p|\big(\Psi_{N-1}^{(A)}(\boldsymbol{x}'_{N-1}), \widetilde\Lambda_N^\dagger(x'_1)\boldsymbol{\Lambda}_N(x_1)\Psi_{N-1}^{(\mathbb{B})}({p,\boldsymbol{x}_{N-1}})\big).
\end{gather*}
The product of layer operators can be factorized into the product of four Baxter operators
\begin{gather}\label{lambdalambda}
\widetilde\Lambda_N^\dagger(x')\boldsymbol{\Lambda}_N(x) = \lambda_N(x,x') Q_{N-1}^\dagger(x) Q_{N-1}^\dagger(-x) \big(Q_{N-1}^\dagger(-x')\big)^{-1} \boldsymbol{Q}_{N-1}(x') ,
\end{gather}
where
\begin{gather*}
\lambda_N(x,x')=\big((-1)^{[2s]}r(x')\big)^{N-1} (-1)^{[s-ix']} q(\pm x,x').
\end{gather*}
Here and below we use a shorthand notation $f(\pm x)\!=\!f(x) f(-x)$, $f(\pm x\pm y)\!=\!f(\pm x+y) f(\pm x-y)$, etc. The diagrammatic representations for the operator $Q_{N}(x)$, $Q_{N}^{-1}(x)$ and $\boldsymbol Q_{N}(x')$ are shown in Figs.~\ref{fig:QX} and~\ref{fig:Blayer}, respectively\footnote{The operator $Q_{N}(x)$ coincides with the operator $Q_{N}^{z_0}(x)$ constructed in~\cite{Derkachov:2016ucn} for $z_0=0$.}. In order to verify that the operator $Q_{N}^{-1}(x)$ defined in such a way is indeed an inverse of $Q_{N}(x)$ one need to check that $Q_{N}(x) Q_{N}^{-1}(x)=\II$. This relation follows immediately from equation~\eqref{delta}.

The proof of the factorization formula~\eqref{lambdalambda} is a bit more complicated. First, using the transformation rules in Appendix~\ref{sect:Diagram} it is easy to show that
\begin{gather*}
\widetilde\Lambda_N^\dagger(x')\boldsymbol{\Lambda}_N(x) = a(x,x') Q_{N-1}^\dagger(x) Q_{N-1}^\dagger(-x) \mathbb{F}(x') .
\end{gather*}
Second, it is not hard to check that at $x'\to x$ the product of the operators takes the form\footnote{The factor $a(x,x')$ is singular in this limit.}
\begin{gather*}
Q_{N-1}^\dagger(x) Q_{N-1}^\dagger(-x) \mathbb{F}(x')|_{x'\to x}\sim Q_{N-1}^\dagger(-x)\boldsymbol Q_{N-1}(x) .
\end{gather*}
It implies that $\mathbb{F}(x)\sim Q_{N-1}^\dagger(-x)\boldsymbol Q_{N-1}(x)$ and results in equation~\eqref{lambdalambda}.

The operators $Q_N(x),\ \boldsymbol Q_N(x')$ satisfy the following exchange relations with the layer operators $\widetilde\Lambda_k$, $\boldsymbol\Lambda_k$
\begin{gather}\label{QLambda}
Q_{k}(x) \widetilde \Lambda_{k}(x')=q(x',x) \widetilde \Lambda_{k}(x')Q_{k-1}(x) ,\\
\boldsymbol Q_{k}(x') \boldsymbol \Lambda_{k}(x)=b(x,x') \boldsymbol \Lambda_{k}(x)\boldsymbol Q_{k-1}(x') .\label{QLambda1}
\end{gather}
The factor $q(x',x)$ is defined in equation~\eqref{q-f-def} and
\begin{gather*}
b(x,x')=\pi^2 a(1-i(x'\pm x)) \frac{ a(s\pm i x)}{a^2(s-ix')} .
\end{gather*}
Again, the identities~\eqref{lambdalambda} and \eqref{QLambda}, \eqref{QLambda1} can be checked diagrammatically by representing the l.h.s.\ and
r.h.s.\ of the identities in a form of Feynman diagrams and mapping one into another with the help of transformation rules given in Appendix~\ref{sect:Diagram}.

Taking into account equations~\eqref{reprPsiA} and \eqref{PsiBopen} one immediately concludes that the functions $\Psi_{N-1}^{(A)}(\boldsymbol{x}'_{N-1})$ and $\Psi_{N-1}^{(\mathbb{B})}({p,\boldsymbol{x}_{N-1}})$ are the eigenfunctions the operators $Q_{N-1}(x)$ and $\boldsymbol Q_{N-1}(x')$, respectively. The corresponding eigenvalues are $\prod\limits_{k=2}^{N}q(x'_k,x_1)$ for the function $\Psi_{N-1}^{(A)}(\boldsymbol{x}'_{N-1})$ and
\begin{gather*}
i^{[-s+ix_1]} \pi a(1 - s + ix_1)[p]^{-s+ix_1}\prod_{k=2}^{N}b(x_k,x'_1)
\end{gather*}
for the function $\Psi_{N-1}^{(\mathbb{B})}({p,\boldsymbol{x}_{N-1}})$. Thus making use of the representation~\eqref{lambdalambda} one can reduce the $N$-point scalar product to the $N-1$-point scalar product. For $N=1$ the scalar product can be easily calculated and, therefore, the answer for general $N$ is obtained by a recursion. We derive in this way
\begin{gather}\label{scAB}
\big(\Psi_N^{(A)}(\boldsymbol{x}'),\Psi_N^{(\mathbb{B})}({p,\boldsymbol{x}})\big)=|p|^{N-1} [p]^{-A_{{X'}}} S_{A} (\boldsymbol{x},\boldsymbol{x}').
\end{gather}
Here $A_{X'}=\sum\limits_{k=1}^N (s-ix'_k)$ and
\begin{gather}\label{SAB}
S_{A} (\boldsymbol{x},\boldsymbol{x}')=\sigma_N i^{[A_{X'}]} \frac{ \prod\limits_{j=1}^{N-1} a^N(s\pm ix_j)}{\prod\limits_{k=1}^N {a^N(s-ix'_k)}}\frac{\prod\limits_{j=1}^{N-1}\prod\limits_{k=1}^N {a(1-i(x'_k\pm x_j))}}{\prod\limits_{k<j} {a(1-i(x'_k+x'_j))}} ,
\end{gather}
where the normalization factor $\sigma_N$ is
\begin{gather*}
\sigma_N=(-1)^{[s]N(N-1)}\pi^{\frac{N(3N-1)}2} .
\end{gather*}
Note that this expression possesses all necessary symmetries: it is symmetric under any permutation of $\boldsymbol{x}$ ($\boldsymbol{x}'$) variables and invariant under the reflections $x_k\to -x_k$, $k=1,\ldots,N-1$. We also want to stress that the $p$-dependence of the scalar product is fixed by the quantum numbers of the eigenstates and can be determined without calculations.

The calculation of the scalar product between the eigenfunctions of $B$ operators goes along the same lines. We omit the details and present the final answer only
\begin{gather*}
\big(\Psi_N^{(B)}(p',\boldsymbol{x}'),\Psi_N^{(\mathbb{B})}({p,\boldsymbol{x}})\big)=\pi \delta^{(2)}(\vec{p}-\vec{p}\,{}') S_{B} (\boldsymbol{x},\boldsymbol{x}') ,
\end{gather*}
where
\begin{gather}\label{SBB}
S_{B} (\boldsymbol{x},\boldsymbol{x}') = \sigma_N \prod_{k=1}^{N-1} \frac{a^N(s\pm ix_k)}{a^N(s-ix'_k)}\frac{\prod\limits_{j,k=1}^{N-1} {a(1-i(x'_k\pm x_j))}}{ \prod\limits_{k<j} {a(1-i(x'_k+x'_j))} } .
\end{gather}
In next section we consider integrals involving these scalar products. Let us discuss analytic properties of the functions~\eqref{SAB}, \eqref{SBB}. First we note that if $\bar x=1-x^*$ then $|a(x)|=1$. It means that all factors like $a(s\pm i x_k)$ are pure phases. Second, it is easy to check that
\begin{gather*}
\biggl|\prod_{k<j}\frac1 {a(1-i(x'_k+x'_j))}\biggr|^2 = \prod_{k<j}[x'_k+x'_j] ,
\end{gather*}
and, hence, this factor is integrable. Thus the only singularities one has to take care of are due to factors $a(1-i(x'_k\pm x_j))$. The function $a(1-ix)$ becomes singular for $x=0$ ($x=-in/2+\nu$) only. Indeed,
\begin{gather*}
a(1-ix) =\frac{\Gamma\big(i\nu-\frac n2\big)}{\Gamma\big(1-i\nu-\frac n2\big)}=\frac{(-1)^{n}\Gamma\big(i\nu+\frac n2\big)}{\Gamma\big(1-i\nu+\frac n2\big)} ,
\end{gather*}
thus the function $a(1-i(x'\pm x))$ is singular only when $n'=\pm n$ and $\nu'=\pm \nu$. In the calculation of the scalar product the divergencies at $x'_k=\pm x_j$ comes from the chain integration. To make this integral finite one can give the variables~$\nu'_k$ a small negative imaginary part, $\operatorname{Im} \nu'_k <0$. This effectively is equivalent to the following $\epsilon$-prescription for avoiding the singularity
\begin{gather*}
{\Gamma(i(\nu'_k \pm \nu_j))}\mapsto {\Gamma(i(\nu'_k\pm\nu_j)+\epsilon)} .
\end{gather*}
In what follows such a prescription of bypassing the po\-les will always be implied.

\section{Integral identities}\label{sect:identities}
In order to obtain the $\mathrm{SL}(2,\mathbb{C})$ integral identities let us re-expand the matrix element $T_{z_0}(\boldsymbol{x},\boldsymbol{x}^\prime)$, equation~\eqref{Txx-result}, over the eigenstates of the operator $\mathbb{B}_N$. Taking into account that
\begin{gather*}
T_{z_0}\Psi_N^{(\mathbb{B})}(p,\boldsymbol x)= e^{-i(pz_0+\bar p\bar z_0)} \Psi_N^{(\mathbb{B})}(p,\boldsymbol x)
\end{gather*}
one gets the following relation
\begin{gather}\label{T-int}
T_{z_0}(\boldsymbol{x},\boldsymbol{x}^\prime)=\int\frac{{\rm d}^2 p}{p \bar p} [p]^{A_x-A_x'} e^{-i(pz_0+\bar p\bar z_0)} \int_{S_I} \mathcal{D}_{N-1} \boldsymbol{u} \boldsymbol{\mu}^{(\mathbb{B})}_N(\boldsymbol{u}) S_{A}(\boldsymbol{u},\boldsymbol{x}') \overline{S_{A}(\boldsymbol{u},\boldsymbol{x})}.
\end{gather}
In order to present the result in a compact form we redefine the variables: $z_k=-ix_k$, $z_{N+k}=ix'_k$ for $k=1,\ldots,N$ ($z_k=m_k/2 + \mu_k$, $\bar z_k=-m_k/2+\mu_k$) and introduce the function~\cite{GelfandGraevRetakh04}
\begin{gather*}
{\boldsymbol{\Gamma}}(z) \equiv {\boldsymbol{\Gamma}}(z,\bar z)=a(1-\bar z)={\Gamma( z)}/{\Gamma(1-\bar z)}
=\frac{\Gamma(m/2+\mu)}{\Gamma(1+m/2-\mu)}=(-1)^m\frac{\Gamma(-m/2+\mu)}{\Gamma(1-m/2-\mu)}.
\end{gather*}
The relation \eqref{T-int} can be written in the form
\begin{gather}
\frac1{2^{N-1}(N-1)!}\prod_{l=1}^{N-1} \sum_{n_l=-\infty}^{\infty} \int_{-i\infty}^{i\infty}\frac{{\rm d}\nu_l}{2\pi i}
\frac{ \prod\limits_{k=1}^{N-1} \prod\limits_{j=1}^{2N} \boldsymbol{\Gamma}(z_j\pm u_k)}
{\prod\limits_{k=1}^{N-1}{\boldsymbol{\Gamma}}(\pm 2u_k) \prod\limits_{1\leq k<j\leq N-1} {\boldsymbol{\Gamma}} (\pm u_k\pm u_j)}\nonumber\\
\qquad{} = \frac{\prod\limits_{1\leq j<k\leq 2N}{\boldsymbol{\Gamma}}(z_j + z_k)}{{\boldsymbol{\Gamma}}\left(\sum\limits_{k=1}^{2N} z_k\right)}.\label{gustafson2}
\end{gather}
Here $u_k\equiv n_k/2+\nu_k$ and the sum goes over all integers or half-integers. All parame\-ters~$n_k$,~$m_k$ are integer or half-integer simultaneously. We assume that $\operatorname{Re} (\mu_k)>0$ so that the series of poles due to ${\boldsymbol{\Gamma}} (z_j-u_k)$ and ${\boldsymbol{\Gamma}} (z_j+u_k)$ lay in the right and left complex half-planes, respectively. This assumption ($\operatorname{Re} (\mu_k)>0$) can be replaced by a~requirement for the series of the poles to be separated by the integration contour. The contour is pinched by the poles whenever the condition
\begin{gather*}
\mu_i+\mu_k=-|m_i+m_k|-{p_{ik}},\qquad p_{ik}\geq 0
\end{gather*}
is fulfilled. The corresponding singularities show up as the poles of ${\boldsymbol{\Gamma}}(z_i+z_k)$ functions in the r.h.s.\ of equation~\eqref{gustafson2}.

The convergence of the integrals and the sum in~\eqref{gustafson2} for large $u_k$ depends on the variables~$z_k$. The integral and sum in~\eqref{gustafson2} con\-ver\-ge absolutely provided that $\operatorname{Re} \left(\sum\limits_{k=1}^{2N} \mu_k\right)<1$ and define an analytic function of~$\mu_k$. Note also that the dependence on the spin $s$ of the spin chain disappears completely.

Next, writing down the functions $\Psi_N^{(A)}(\boldsymbol{x}')$ and $\Psi_N^{(\mathbb{B}}(x)$ in the basis provided by the eigenstates of the operator $B_N$ one gets
\begin{gather*}
S_A(\boldsymbol x, \boldsymbol x') = \pi \int \mathcal{D}_{N-1}\boldsymbol u \boldsymbol \mu_N^{(B)}(\boldsymbol u)
S_B(\boldsymbol x, \boldsymbol u) \overline{S_{BA}(\boldsymbol x',\boldsymbol u)}.
\end{gather*}
Using the explicit expressions~\eqref{scAB}, \eqref{SAB}, \eqref{SBB} for the functions $S_{A,(B,AB)}$ and introducing the notations $z_k=ix'_k =m_k/2+\mu_k$, $w_k=-ix_k= p_k/2+\xi_k$ we obtain
\begin{gather}
\frac1{(N-1)!}\prod_{l=1}^{N-1} \sum_{n_l=-\infty}^{\infty} \int_{-i\infty}^{i\infty}\frac{{\rm d}\nu_l}{2\pi i}
\frac{\prod\limits_{k=1}^{N-1}\left( \prod\limits_{j=1}^{N-1}\boldsymbol{\Gamma}(u_k\pm w_j)\prod\limits_{m=1}^N \boldsymbol{\Gamma}(z_m-u_k)\right)}{
\prod\limits_{1\leq m<j\leq N-1} \boldsymbol{\Gamma}(u_m\pm u_j)\boldsymbol{\Gamma}(u_j- u_m)}\nonumber\\
\qquad{} =\frac{\prod\limits_{k=1}^{N}\prod\limits_{j=1}^{N-1}\boldsymbol{\Gamma}(z_k\pm w_j)}{\prod\limits_{1\leq k<m\leq N} \boldsymbol{\Gamma}(z_m+ z_k)}.\label{gustafson3}
\end{gather}
Here, as in equation~\eqref{gustafson2}, $u_k\equiv n_k/2+\nu_k$. All para\-me\-ters $\{m_k,p_j,n_i\}$ are integer or half-integer numbers si\-mul\-taneously. The integration contours separate the poles the functions $\boldsymbol{\Gamma}(z_m-u_k)$ and $\boldsymbol{\Gamma}(u_k\pm w_j)$. The singularities due to the $\boldsymbol{\Gamma}$ functions in the numerator in the r.h.s.\ of~\eqref{gustafson3} come from pinching of the integration contours. The singularities coming from the function $\boldsymbol{\Gamma}(z_m+z_k)$ in the denominator can be traced to the divergence of the integral in the regime $u_m\sim -u_k \to \infty$.

We also present here the $\mathrm{SL}(2,\mathbb{C})$ analog of the first Gustafson integral. It was derived in~\cite{Derkachov:2016ucn} and can be written as
\begin{gather}
\frac1{(N-1)!}\prod_{l=1}^{N-1} \sum_{n_l=-\infty}^{\infty} \int_{-i\infty}^{i\infty}\frac{{\rm d}\nu_l}{2\pi i}
\frac{\prod\limits_{j=1}^{N}\prod\limits_{k=1}^{N-1}\boldsymbol{\Gamma}(z_j-u_k)\boldsymbol{\Gamma}(u_k+w_j)}{
\prod\limits_{1\leq m<j\leq N-1} \boldsymbol{\Gamma}(u_m- u_j)\boldsymbol{\Gamma}(u_j- u_m)}\nonumber\\
\qquad{} =\frac{\prod\limits_{k,j=1}^{N}\boldsymbol{\Gamma}(z_k+ w_j)}{ \boldsymbol{\Gamma}\left(\sum\limits_{k=1}^N (z_k+w_k)
\right)}.\label{gustafson1}
\end{gather}
As in the previous case the integration contours have to separate the series of poles of $\boldsymbol{\Gamma}$-functions in the numerator. The integral and sum converge absolutely provided that
\begin{gather*}
\operatorname{Re} \left(\sum_{k=1}^{N}(\mu_k+\xi_k)\right)<1.
\end{gather*}

For this integral there is no difference between integer and half-integer cases, since the equation is invariant under the transformation $u_k\mapsto u_k+1/2$, $z_k\mapsto z_k+1/2$ and $w_k\mapsto w_k-1/2$. Note that this is not true for other two integrals.

Below we write down integrals~\eqref{gustafson1} and \eqref{gustafson2} for $N=2$ in the explicit form\footnote{For $N=2$ the integral~\eqref{gustafson3} is a partial case of~\eqref{gustafson1}.}. We get
\begin{gather}
\sum_n\int\frac{{\rm d}\nu}{2\pi i} \prod_{k=1,2} \frac{\Gamma\big(z_k-\frac n2 -\nu\big)\Gamma\big(w_k+\frac n2 +\nu\big)}{\Gamma\big(1-\bar z_k-\frac n2 +\nu\big)\Gamma\big(1-\bar w_k+\frac n2 -\nu\big)}\nonumber\\
\qquad{} =\frac{\Gamma\big(1-\sum_k(\bar z_k+\bar w_k)\big)}{\Gamma\big(\sum_k (z_k+w_k)\big)}\prod_{k,j=1,2} \frac{\Gamma(z_k+w_j)}{\Gamma(1-\bar z_k-\bar w_j)}\label{closedN1}
\end{gather}
for the integral~\eqref{gustafson1} and
\begin{gather}
\frac12 \sum_n \int\frac{{\rm d}\nu}{2\pi i} (-1)^{2n} \big(n^2-4\nu^2\big)
\prod_{k=1}^4\frac{\Gamma \big(z_k\pm \big(\frac{n}2+ \nu\big)\big)}{\Gamma\big(1-\bar z_k\pm \big(\frac{n}2 - \nu\big)\big)}\nonumber\\
\qquad{} =\frac{\Gamma\big(1-\sum_k\bar z_k\big)}{\Gamma\big(\sum_k z_k\big)}\prod_{k<j} \frac{\Gamma(z_k+z_j)}{\Gamma(1-\bar z_k-\bar z_j)}\label{openN1}
\end{gather}
for the integral~\eqref{gustafson2}. Again, the sum goes over integer or half-integer $n$.

The integrals~\eqref{closedN1} and \eqref{openN1} are the generalization of Barnes' first lemma and de Branges--Wilson in\-te\-gral~\cite{MR0417337,MR579561} to the complex case.

\section{Summary}\label{sect:summary}

We have derived a generalization of Gustafson's integrals to the complex case. The integrals have the same functional form as in the real case save that the $\Gamma$-func\-tions have to be replaced by the $\boldsymbol \Gamma$-functions associated with the complex field $\mathbb{C}$~\cite{GelfandGraevRetakh04} and the integration mea\-sure have to be appropriately modified.

Our analysis relies on the QISM technique, espe\-cially the separation of variables method. For the $\mathrm{SL}(2,\mathbb{C})$ spin chain the eigenfunctions of elements of the monodromy matrix are known explicitly. Using the diagrammatic approach we calculated some scalar products between the corresponding eigen\-functions of the closed and open spin chains. The sca\-lar products are given by a product of the $\boldsymbol\Gamma$-functions associated with the complex field~$\mathbb{C}$. Using the com\-ple\-te\-ness of the SoV representation and expanding vectors in the scalar products over an appropriate basis we obtain generalization of Gustafson's integrals to the complex case.

We expect that the same technique can be applied to trigonometric and elliptic spin chains and gives rise to new $q$-beta and elliptic Gustafson type integrals.

\appendix

\section{The diagram technique}\label{sect:Diagram}

Throughout this paper we used a diagrammatic representation for the kernels of relevant ope\-ra\-tors. Relations between ope\-rators are equivalent to the corresponding relations between ope\-ra\-tor's kernels. The most convenient way to check such relations is to prove the equivalence of the corresponding diagrams (kernels). It can be done diagrammatically with the help of several simple identities~-- integration rules. Below we give some of these rules (see also~\cite{Derkachov:2001yn}).

\medskip

\noindent
$\bullet$ An arrow with the index $\alpha$ directed from $w$ to~$z$ stands for a propagator $D_\alpha(z-w)=[z-w]^{-\alpha}$:
\begin{gather*}
\includegraphics[width=0.35\linewidth]{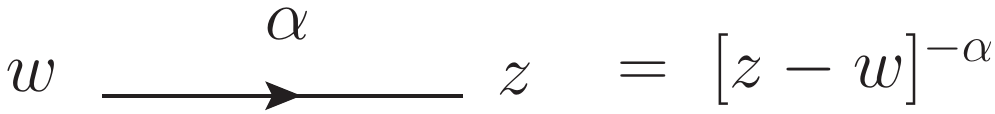}
\end{gather*}
It has the following properties
\begin{gather*}
D_\alpha(z-w)=(-1)^{[\alpha]} D_\alpha(w-z).
\end{gather*}

\noindent
$\bullet$ The Fourier transform reads
\begin{gather*}
\int {\rm d}^2 z e^{i(pz+\bar p\bar z)}D_\alpha(z)=\pi i^{\alpha-\bar\alpha} a(\alpha) D_{1-\alpha}(p) ,
\end{gather*}
where the function $a$ is
\begin{gather*}
a(\alpha)\equiv a(\alpha,\bar\alpha)={\Gamma(1-\bar\alpha)}/{\Gamma(\alpha)} ,
\end{gather*}
It has the following properties
\begin{alignat*}{3}
& a(\alpha) a(1-\bar\alpha) =1, \qquad && a(\alpha) = -{\alpha\bar\alpha}{ a(1+\alpha)},& \\
& a(\alpha)a(1-\alpha) =(-1)^{[\alpha]},\qquad && a(\alpha)=(-1)^{[\alpha]} a(\bar\alpha).&
\end{alignat*}

\noindent
$\bullet$ Chain rule
\begin{gather*}
\int\frac{ {\rm d}^2 w}{[z_1-w]^\alpha [w-z_2]^{\beta}}=\pi \frac{ a(\alpha)a(\beta)}{a(\gamma)}\frac{1}{[z_1-z_2]^{\gamma}} ,
\end{gather*}
where $\gamma=\alpha+\beta-1$. Its diagrammatic form is
\begin{gather*}
\includegraphics[width=0.45\linewidth]{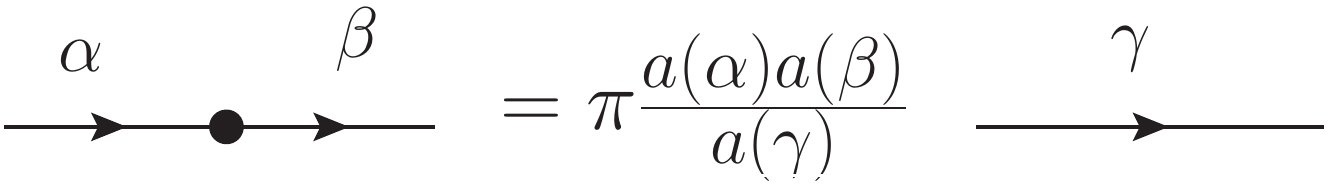}
\end{gather*}
For a special case $\beta\to 2-\alpha$ one gets
\begin{gather}\label{delta}
\int\frac{{\rm d}^2 w}{[z_1-w]^\alpha [w-z_2]^{2-\alpha}}=\pi^2a(\alpha,2-\alpha)\delta^{(2)}(z_1-z_2).
\end{gather}

\noindent
$\bullet $ Star-triangle relation
\begin{gather*}
\includegraphics[width=0.5\linewidth]{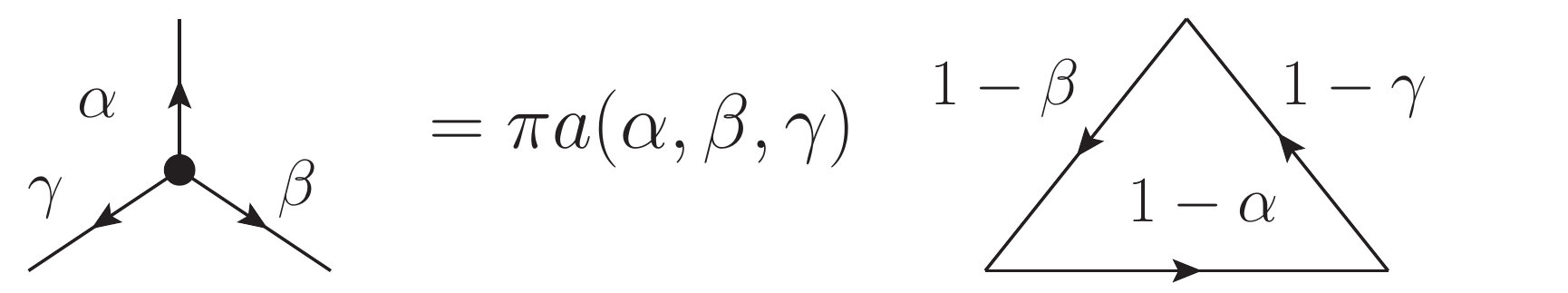}
\end{gather*}

\noindent
$\bullet$ Cross relation
\begin{gather*}
\includegraphics[width=0.7\linewidth]{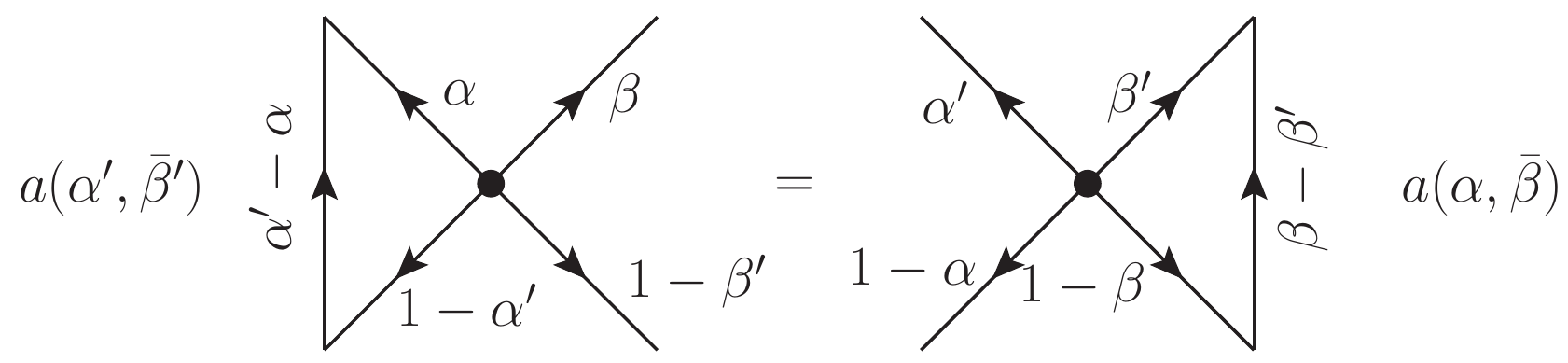}
\end{gather*}
where $\alpha+\beta=\alpha'+\beta'$.

\subsection*{Acknowledgements}
This study was supported by the Russian Science Foun\-da\-tion project $\text{N}^{\text{o}}$ 14-11-00598 and Deutsche Forschungsgemeinschaft (A.~M.), grant MO~1801/1-2.

\pdfbookmark[1]{References}{ref}
\LastPageEnding

\end{document}